\documentclass[prc,aps,amsmath,amssymb,letter]{revtex4}
\usepackage{graphicx}
\usepackage{epsfig,rotate,latexsym}
\usepackage{hyperref}
\usepackage{rotating}
\usepackage{subfig}
\usepackage{amsmath}
\usepackage{amssymb}
\usepackage{mathtools}
\begin{document}
\begin{flushright}
\end{flushright}
\begin{center}
{\Large\bf  Transport coefficients of hadronic matter in a van der Walls 
hadron resonance gas model}\\[1cm]
{\large Ranjita K. Mohapatra$^{1}$, Hiranmaya Mishra$^{2}$, Sadhana Dash$^{1}$ and Basanta K. Nandi$^{1}$  }\\[0.5cm]
{\it $^1$Department of Physics, Indian Institute of Technology Bombay,
Mumbai 400076, India 
 }\\
 {\it $^2$Theory Division, Physical Research Laboratory, Navrangpura,
Ahmedabad 380009, India}\\[1cm]
\end{center}

\begin{abstract}

We estimate the transport coefficients like shear and bulk viscosities of hot hadronic matter
in van der Walls hadron resonance gas ( VDW HRG ) model in the relaxation time approximation.
We also have compared these results with excluded volume hadron resonance ( EV HRG ) 
calculations. $\eta/s$ decreases as the temperature of the hadronic system increases at a
fixed baryon chemical potential. $\eta/s$ in VDW HRG is always less than that of EV HRG case 
due to the large entropy density in VDW HRG compared to EV HRG case. At a fixed 
chemical potential, $\zeta/s$ in VDW HRG is also less than that of EV HRG case. We also 
have estimated these transport coefficients along the freezeout curve. There is an
increase in chemical freezeout temperature in VDW HRG case determined from the universal 
condition $E/N =\epsilon/n \sim 1$ $GeV$. We also have calculated the variation of attraction 
parameter along the freezeout curve $E/N =\epsilon/n \sim 1$ $GeV$ keeping the 
freezeout parameters same as in ideal HRG and the repulsion parameter fixed. We observe
a nontrivial variation of attraction parameter along the freezeout curve where it increases
in the meson dominated region and decreases in the baryon dominated region along the chemical
freezeout curve. $\eta/s$ in EV HRG is always large than that VDW HRG along the freezeout curve. 
This is also true for $\zeta/s$.
     
\end{abstract}

\maketitle

\section{INTRODUCTION}

The large value of elliptic flow observed at Relativistic Heavy
Ion Collider (RHIC) and Large Hadron Collider (LHC) confirms that 
there is a strongly interacting quark gluon plasma (sQGP) produced
at these collision energies \cite{paul,alice}. The "almost perfect fluid" nature of QGP is 
produced at RHIC energies with shear viscosity to entropy density ratio i.e
$\eta/s\leq 0.2$. This value is close to anti-de Sitter/conformal 
field theory (AdS/CFT) lower bound on $\eta/s$ famously known as KSS (Kovtun-Son-Starinets)
bound i.e $\eta/s\geq 1/{4\pi}$ \cite{son}. According 
to the AdS/CFT, for any strongly interacting fluid $\eta/s$ has a lower bound
i.e  $\eta/s\geq 1/{4\pi}$. This bound has been verified for different 
fluids like $H_{2}O$, $N_{2}$ and $He_{2}$ etc \cite{lacey}. The bulk viscosity
$\zeta$ is zero for any fluid which has conformal symmetry. However, QCD does
not have the conformal symmetry around the critical temperature because the trace 
anamoly ${(\epsilon-3P)}/{T^4}$ shows a peak as observed in the lattice simulations
\cite{bazav,bor}. Perturbative QCD calculations have shown a much higher value of the 
shear viscosity to the entropy density ratio i.e $\eta/s\geq 1$ \cite{arnold}. It has been 
observed that the bulk viscosity is 1000 times smaller than the shear 
viscosity in the perturbative QCD \cite{arnold}. But, the non-perturbative nature 
of QCD is relevant around the phase transition or the crossover region which        
shows the bulk viscosity is not very small compared to the shear viscosity.
The shear viscosity of QGP decreases due to nonzero bulk viscosity coefficient.
The nonzero value of the bulk viscosity with nonzero shear viscosity explains 
the experimental data on the multiplicity, the average transverse momentum and the elliptic 
flow coefficients very well \cite{prl2015}.

The transport coefficients like the shear and the bulk viscosities govern the 
non-equilibrium system towards the equilibrium state. The shear viscosity shows the 
resistance to any deformation in the system due to
the shear stress and the bulk viscosity shows the resistance to change in the volume
of the system. It has been observed that the ratio of shear viscosity to entropy density
shows a minimum around the phase transition region \cite{csernai} and the ratio of bulk viscosity to
entropy density shows a maximum around this region \cite{kapusta1,kharz}.
So, these transport coefficients are very important to study the phase 
transition region and also the QCD critical point.   

The transport coefficients like the shear and the bulk viscosities of the QGP phase 
and the hadronic matter have been calculated by various methods. In principle, the transport 
coefficients can be calculated directly from QCD using Kubo formulas \cite{tuchin,plumari}.
However, these calculations are very difficult due to strongly interacting QCD 
in the region of interest to us. The transport coefficients also have been 
calculated using Boltzmann equation
in the relaxation time approximation \cite{gavin,hm}. In this relaxation time approximation, 
the collision integral in Boltzmann equation has been approximated in such a way 
that the distribution function reaches to the equilibrium distribution function 
exponentially with a relaxation time $\tau$. These transport coefficients also have been 
calculated in Chapmann-Enskog method \cite{groot,plumari}. These also have been
calculated using effective field theory models and quasiparticle model 
\cite{marty,hm2,kapusta2}. These are also other kinetic  
theory calculations on transport coefficients \cite{bass,muronga,dubado,chen,itakura,sasaki} etc. 

In this work, we would like to estimate the shear and the bulk viscosity of the
hot hadronic matter in a van der Walls hadron resonance gas (VDW HRG) using the 
relaxation time approximation. The ratio of shear viscosity to entropy density 
of hadronic matter also has been shown in
VDW HRG with different attraction and repulsion parameters than described here 
in ref\cite{nachiketa}. 
The ideal HRG (IHRG) model which describes the hadrons as point particles successfully
explains the lattice data for different thermodynamic quantities at low temperature.
But, there is a mismatch between the lattice results and IHRG calculations
around the critical temperature \cite{munzi,stachel}. This mismatch has been taken 
care of upto some level by taking finite radius of hadrons in the excluded 
volume (EV HRG) procedure in a grand canonical 
ensemble \cite{rischke}. There is repulsive interaction interaction between hadrons due to finite 
radius ( hard core repulsion) in EV HRG model. However, apart from hard core 
repulsion, there is van der Walls attraction between the hadrons ( VDW HRG model). 
The calculations of the thermodynamic quantities in VDW HRG model matches well with 
the lattice results around the critical temperature \cite{vdw1,vdw2,vdw3,vdw4}. 

This paper is organized as follows.
we discuss the essential aspects of the HRG model in section II. 
Section III describes the estimation of the transport coefficients like shear and bulk 
viscosities in the relaxation time approximation. Section IV describes 
the results obtained in VDW HRG model. Then we conclude in Section V.  

\section{HADRON RESONANCE GAS MODEL}

\subsection{ Ideal and excluded volume model}

The IHRG model successfully describes the thermodynamic quantities and matches with 
the lattice QCD calculations. The hadrons are treated as point particles in the model and the 
thermodynamic quantities blows up as the system approaches the critical temperature.
However, the hadrons are not point particles, there is repulsive interaction 
between these hadrons due to their finite size and this has been taken in EV HRG model.

The grand canonical partition function of IHRG for each hadron species is written as

\begin{equation}
{\ln Z^{id}_{i}}=\pm {Vg_{i}\int{\frac{d^3p}{(2\pi)^3}\ln\left[1\pm{e^{-(E_{i}-\mu_{i})/T}}\right]}}\end{equation}
  
Here $\pm$ corresponds to fermions and bosons respectively. Here V is the volume of
the system, $g_{i}$ is the spin degeneracy factor, $E_{i}=\sqrt{p^2+{m_{i}}^2}$ is the single
particle energy and $\mu_{i}=B_{i}\mu_{B}+S_{i}\mu_{S}+Q_{i}\mu_{Q}$ is the
chemical potential. Here $B_{i}$, $S_{i}$ and $Q_{i}$ are the baryon number, strange
and electric charge of the particle and $\mu_{B}$, $\mu_{S}$ and $\mu_{Q}$ are the
corresponding chemical potentials. All the thermodynamic quantities
like pressure, energy density and entropy density etc. can be derived from this
partition function.

In a thermodynamically consistent EV HRG model, the pressure is given by 

\begin{equation}
P^{EV}{(T, \mu)} = \sum_{i}{p_{i}^{\rm id} (T, \mu^{*})}
\end{equation}

where the chemical potential for ith hadron is given by

\begin{equation}
\mu^{*}=\mu - V_{ex}P^{EV}
\end{equation}

where $V_{ex} = {16\pi r^3}/3$ the excluded volume for each hadron with hard core radius r

The number density, entropy density and energy density in EV HRG is given by

\begin{align}
\label{eq:nbev1}
n^{EV}{(T, \mu)} ~ = ~
\frac {\sum_{i} {n_{i}^{\rm id} (T, \mu_i^{*})}}{1 + V_{ex} 
{\sum_{i} {n_{i}^{\rm id} (T, \mu_i^{*})}}}.\\
\label{eventrp1}
s^{EV}{(T, \mu)} ~ = ~
\frac {\sum_{i} {s_{i}^{\rm id} (T, \mu_i^{*})}}{1 + V_{ex} 
{\sum_{i} {s_{i}^{\rm id} (T, \mu_i^{*})}}}.\\
\epsilon^{EV}{(T, \mu)} ~ = ~
\frac {\sum_{i} {\epsilon_{i}^{\rm id} (T, \mu_i^{*})}}{1 + V_{ex} 
{\sum_{i} {n_{i}^{\rm id} (T, \mu_i^{*})}}}.\\
\end{align}

\subsection{ van der Walls hadron resonance gas}
  
The grand canonical ensemble formulation of the full VDW
equation with both attractive and repulsive interactions, was developed 
in \cite{vdw1,vdw2,vdw3,vdw4}. The attractive parameter $a$ 
and the repulsive parameter $b$ of VDW HRG are uniquely fixed by reproducing the nuclear 
saturation density $n_0 = 0.16$~fm$^{-3}$ and binding energy $E/A = -16$~MeV of the 
ground state of nuclear matter. For nucleons the values $a = 329$~MeV$~$fm$^3$ and
$b = 3.42$~fm$^3$ were obtained. This value of $b$ corresponds to a nucleon 
radius $\sim $0.58 fm from the relation $b = {16\pi r^3}/3$. 
This model predicts a liquid-gas first order phase transition in nuclear matter 
with a critical point at $T_c \simeq 19.7$~MeV 
and $\mu_c \simeq 908$~MeV.

However, the VDW interactions have been taken between baryons-baryons and antibaryons-antibaryons
in VDW HRG model \cite{vdw4}. The VDW parameters $a$ and $b$ for all (anti)baryons 
are assumed to be equal to those of nucleons. The baryon-antibaryon, meson-meson, and meson-
(anti)baryon VDW interactions are neglected. The baryon-antibaryon VDW interactions 
are neglected because short-range interactions between baryons
and antibaryons may be dominated by annihilation processes \cite{vdw4}. The meson-meson VDW 
interactions also have been neglected because of the significant mesonic eigen volume,
comparable to those of baryons, leads to significant suppression of thermodynamic
functions in the crossover region at $\mu_{B} = 0$~ which does not match with lattice
data \cite{vdw4}. The attractive interactions between mesons lead to resonance formation,
which have been already included in the HRG model. 

The VDW HRG model consists of three subsystem: 
Noninteracting mesons, VDW baryons, and VDW antibaryons.
The total pressure is given by

\begin{equation}
P(T, \mu) = P_M(T, \mu) + P_B(T, \mu) + P_{\bar{B}}(T, \mu).
\end{equation}
where
\begin{equation}
P_M(T, \mu) =
\sum_{i \in M} p_{i}^{\rm id} (T, \mu_i). \\
\end{equation}

\begin{equation}
\label{eq:pres1}
P_B(T, \mu) =
\sum_{i \in B} p_{i}^{\rm id} (T, \mu_i^{B*}) - a\,n_B^2. \\
\end{equation}

\begin{equation}
\label{eq:pres2}
P_{\bar{B}}(T,\mu) =
\sum_{i \in \bar{B}} p_{i}^{\rm id} (T, \mu_i^{\bar{B*}}) - a\,n_{\bar{B}}^2.
\end{equation}

Here $M$ stands for mesons, $B$ for baryons, and $\bar{B}$ for antibaryons, 
$p_{i}^{\rm id}$ is the Fermi or Bose ideal gas pressure,
$\mu_i^{B(\bar{B})*} = \mu_i - b\,P_{B(\bar{B})} - a\,b\,n_{B(\bar{B})}^2 + 2\,a\,n_{B(\bar{B})}$,
and $n_B$ and $n_{\bar{B}}$ are total densities of baryons and antibaryons respectively.

The number density, entropy density and energy density for baryons and antibaryons 
in VDW HRG are given by:

\begin{align}
\label{eq:nb1}
n_{B(\bar{B})} ~ = ~
\frac {\sum_{i \in {B(\bar{B})}} {n_{i}^{\rm id} (T, \mu_i^{B(\bar{B})*})}}{1 + b 
\sum_{i \in B(\bar{B})} n_{i}^{\rm id} (T, \mu_i^{B(\bar{B})*})}.\\
\label{entrp1}
s_{B(\bar{B})}
~ =~
\frac{ \sum_{i \in B(\bar{B})} s_{i}^{\rm id} (T, \mu_i^{B(\bar{B})*})}{1 + b 
\sum_{i \in B(\bar{B})} n_{i}^{\rm id} (T, \mu_i^{B(\bar{B})*})}.\\
\epsilon_{B(\bar{B})}
~ =~
\frac{ \sum_{i \in B(\bar{B})} \epsilon_{i}^{\rm id} (T, \mu_i^{B(\bar{B})*})}{1 + b 
\sum_{i \in B(\bar{B})} n_{i}^{\rm id} (T, \mu_i^{B(\bar{B})*})}  
- a\,n_{B(\bar{B})}^2(T,\mu).\\
\label{edens}
\end{align}

The thermodynamic quantities in VDW HRG model obey the self consistency relation
$Ts=\epsilon + P - \mu n$. 
We have incorporated all the hadrons listed in the particle data book 
upto mass 3 GeV \cite{pdg2012}. 

\section {TRANSPORT COEFFICIENTS IN RELAXATION TIME APPROXIMATION}

Here we briefly describe the estimation of transport coefficients like shear and bulk 
viscosity in the relaxation time approximation \cite{gavin}.    

The Boltzmann transport equation of kinetic theory is given by

\begin{equation}
\frac{\partial f_{p}}{\partial t}+ v_{p}^{i}\frac{\partial f_{p}}{\partial x^{i}}=C[f_{p}].
\label{boltz}
\end{equation}   

where $f_p$ is the single particle distribution function, $\vec v_{p}=\vec p/E_{p}$ is 
particle velocity and $C[f_p]$ is the collision integral. The collision integral 
gives the rate of change of the distribution function
due to collisions of the constituent particles of the system. However, this is 
a very complicated integral to solve. So, this collision integral has been 
approximated such that the non-equilibrium distribution function $f_p$ approaches to
the equilibrium distribution ${f_p}^0$ exponentially with a relaxation time $\tau$
which is of the order of collision time. Hence, the collision integral in the
relaxation time approximation is given by

\begin{equation}
C[f_{p}]\backsimeq -\frac{(f_{p}-f_{p}^{0})}{\tau(E_{p})} = -\frac{\delta f_{p}}{\tau}
\label{rtacoll}
\end{equation}

The relaxation time $\tau$ depends on the energy of the particle.
The equilibrium distribution function ${f_p}^0$ is given by

\begin{equation}
f_{p}^{0}=\frac{1}{exp\bigg(\frac{E_{p}-\vec p.\vec u-\mu}{T}\bigg)\pm1}
\label{eqdist}
\end{equation}

 Here $\vec u$ is the fluid velocity and $\pm$ corresponds to fermions and
bosons respectively.

In hydrodynamics, the stress energy tensor is defined as 

\begin{equation}
T^{\mu\nu}=T_{0}^{\mu\nu}+T_{dissi}^{\mu\nu}
\label{stress}
\end{equation}

where $T_{0}^{\mu\nu}$ is the ideal part of the fluid when there is no dissipation of the fluid 
taken into account. $T_{dissi}^{\mu\nu}$ is the dissipative part of the fluid from where 
shear and bulk viscosities can be calculated.

The shear and the bulk viscosities are related to the dissipative part of stress energy
tensor

\begin{equation}
T_{dissi}^{ij}=-\eta\bigg(\frac{\partial u^{i}}{\partial x^{j}}+\frac{\partial u^{j}}
{\partial x^{i}}\bigg)-(\zeta-\frac{2}{3}\eta)\frac{\partial u^{i}}{\partial x^{j}}\delta^{ij}
\label{dissi}
\end{equation}

The stress energy tensor in terms of the distribution function given by
\begin{equation}
T^{\mu\nu}=g \int \frac{d^{3}p}{(2\pi)^{3}{p_0}}p^{\mu}p^{\nu}f_{p}
\label{stressdist}
\end{equation}

Here g is the degeneracy of the particle.
Also
\begin{equation}
f_p={f_p}^0+\delta f_p
\label{dist}
\end{equation}

So using Eq. (\ref{stressdist}) and Eq. (\ref{dist}), the dissipative part of 
stress energy tensor given by 

\begin{equation}
T_{dissi}^{ij}=g\int \frac{d^{3}p}{(2\pi)^{3}{p_0}}p^{i}p^{j}\delta f_{p} 
\label{dissi2}
\end{equation}

From Eqs. (\ref{boltz}) and (\ref{rtacoll})

\begin{equation}
\delta f_{p}=-\tau(E_{p})\bigg(\frac{\partial f_{p}^{0}}{\partial t}+ v_{p}^{i}\frac
{\partial f_{p}^{0}}{\partial x^{i}} \bigg)
\label{deltaf}
\end{equation}

For one dimensional flow of the form $u^{i}=(u_{x}(y),0,0)$,  Eq. (\ref{dissi}) 
simplifies to 

\begin{equation}
T^{xy}_{dissi}=-\eta\partial u_{x}/\partial y. 
\label{txy1}
\end{equation}

Using Eqs. (\ref{dissi2}), (\ref{deltaf}) and (\ref{eqdist}), one can obtain

\begin{equation}
T^{xy}_{dissi}=\bigg\{-\frac{g}{T}\int\frac{d^{3}p}{(2\pi)^{3}}\tau(E_{p})\bigg(\frac{p_{x}p_{y}}{E_{p}}\bigg)^{2}f_{p}^{0}\bigg\}\frac{\partial u_{x}}{\partial y}
\label{txy2}
\end{equation}

Equating Eq. (\ref{txy1}) and Eq. (\ref{txy2}), we obtain the expression for
shear viscosity 

\begin{equation}
\eta=\frac{g}{15T}\int\frac{d^{3}p}{(2\pi)^{3}}\tau(E_{p})\frac{p^{4}}{E_{p}^{2}}f_{p}^{0}
\label{etaf}
\end{equation}

Taking the trace of Eq. (\ref{dissi}), we obtain bulk viscosity

\begin{equation}
(T_{dissi})^{i}_{i}=-3\zeta \frac{\partial u^{i}}{\partial x^{i}}
\label{bulkdissi}
\end{equation}

From Eq. (\ref{dissi2}), we also get

\begin{equation}
(T_{dissi})^{i}_{i}=-g\int\frac{d^{3}p}{(2\pi)^{3}}\tau(E_{p})\frac{p^{2}}{E_{p}}\bigg
(\frac{\partial f_{p}^{0}}{\partial t}+ v_{p}^{i}\frac{\partial f_{p}^{0}}{\partial x^{i}} \bigg)
\label{bulkdissi2}
\end{equation}

From Eqs. (\ref{bulkdissi}) and (\ref{bulkdissi2}), and using the energy momentum 
conservation equation $\partial_{\mu}T^{\mu\nu}=0$, one can obtain

\begin{equation}
\zeta=\frac{g}{T}\int \frac{d^{3}p}{(2\pi)^{3}}\tau(E_{p}) f^{0}_{p}\bigg
[E_{p}C_{n_{B}}^{2} \pm \bigg(\frac{\partial P}{\partial n_{B}}\bigg)_{\varepsilon}
-\frac{p^{2}}{3E_{p}}\bigg]^{2}
\label{zetaf}
\end{equation}

Here $\pm$ corresponds to particles and anti particle respectively. 
${C_{nb}}^2 = \frac{\partial P}{\partial\varepsilon}|_{n_{B}}$, is the speed of sound 
at constant baryon density.

For a hadron resonance gas, the shear and bulk viscosities can be written as a sum over
all the particles included in this model

\begin{equation}
\eta=\frac{1}{15T}\sum_{a}\int \frac{g_ad^{3}p}{(2\pi)^{3}}\frac{{p}^{4}}{E_{a}^{2}}
({\tau}_{a}f_{a}^{0})
\label{shearf}
\end{equation}

\begin{equation}
\zeta = \frac{1}{T}\sum_{a}\int \frac{g_ad^{3}p}{(2\pi)^{3}}\bigg\{\tau_{a} f^{0}_{a}
\bigg[E_{a}C_{n_{B}}^{2} \pm \bigg(\frac{\partial P}{\partial n_{B}}\bigg)_{\varepsilon} 
-\frac{ p^{2}}{3E_{a}}\bigg]^{2}
\label{bulkf}
\end{equation}
 
Here $a$ corresponds to all the hadrons taken in VDW HRG model.

The average relaxation time for a hadron $a$ is given by

\begin{equation}
{\tilde\tau}_{a}^{-1}=\sum_{b}n_{b}\langle\sigma_{ab}v_{ab}\rangle 
\label{taue}
 \end{equation}

here the summation is over all other hadrons in VDW HRG model.
$n_b$ is the number density of the hadron b.

For any scattering process $a(p_{a})+b(p_{b})\rightarrow a(p_{c})+b(p_{d})$, 
the thermal averaged cross section times the relative velocity is given by

\begin{equation}
\langle\sigma_{ab}v_{ab}\rangle=\frac{\sigma}{8Tm_{a}^{2}m_{b}^{2}K_{2}(\frac
{m_{a}}{T})K_{2}(\frac{m_{b}}{T})}\int_{(m_{a}+m_{b})^2}^{\infty}dS
\frac{[S-(m_{a}-m_{b})^{2}]}{\surd S}[S-(m_{a}+m_{b})^{2}]K_{1}(\surd S/T)
\label{sigmaav}
\end{equation}

Here $K_n$ is the modified Bessel function of order n. The cross section $\sigma$ 
in Eq. (\ref{sigmaav}) in terms of hadron radius $r$ is given by $\sigma=4\pi r^2$. 

\section {RESULTS AND DISCUSSION}

In order to calculate the shear and the bulk viscosity of the hot hadronic matter, one needs 
an estimation of the average relaxation time $\tau$ with respect to the temperature of the system.
Here the average is taken over all the hadrons with same radius $\sim 0.5$ fm.
Fig.1 shows the variation of the relaxation time with respect to the temperature at
three different chemical potentials. At a fixed chemical potential, the relaxation 
time decreases as the temperature increases. This is due to the fact that the 
number density increases as the temperature increases and $\tau$ is inversely proportional
to the number density according to Eq. (\ref{taue}). The average relaxation time also
decreases as the chemical potential increases at a fixed temperature. This is because 
the baryon number density increases as the chemical potential increases. At $\mu_{B} = 0$, 
the average relaxation time decreases from $\sim 10$ fm to $\sim 0.2$ fm as the temperature
of the hadronic system increases from 100 MeV to 200 MeV.   
  
\begin{figure*}[!hpt]
\begin{center}
\leavevmode
\epsfysize=6truecm \vbox{\epsfbox{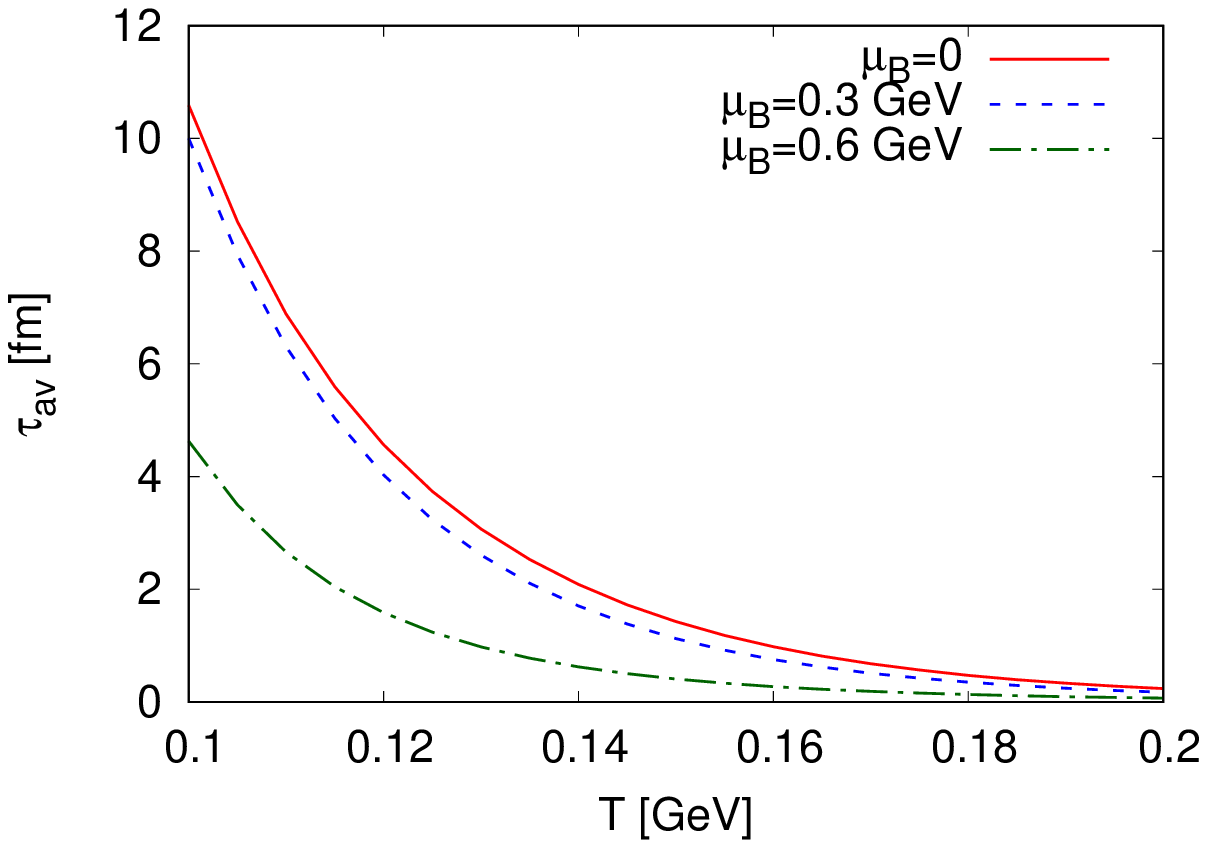}}
\end{center}
\caption{}{The average relaxation time with respect to temperature 
at three different baryon chemical potentials.}
\label{Fig.1}
\end{figure*} 
 
The normalized entropy density for VDW HRG model is given for different conditions in 
Fig.2 at different chemical potentials. At $\mu_{B} = 0$, the solid curve shows 
the variation of normalized entropy density for an ideal HRG model. The dotted 
curve represents the normalized entropy density when the attraction parameter $a$ of 
VDW HRG model is zero, but the repulsion parameter $b = 3.42$~fm$^3$ as in VDW HRG
model. This is equivalent to the EV HRG model where there is only hard core 
repulsion between the hadrons. However, here the repulsion has been taken only
between baryon-baryon and antibaryon-antibaryon (as in VDW HRG model). Due
to only the hard core repulsion i.e $a = 0$, the normalized entropy density decreases
compared to the ideal HRG. This is prominent after a temperature 160 MeV at
$\mu_{B} = 0$. When the repulsion parameter $b$ of VDW HRG is zero ( equivalent to 
point particles as in the ideal HRG), but the attraction term is $a = 329$~MeV$~$fm$^3$ 
as in VDW HRG, the entropy density increases compared to the ideal HRG due to the 
attraction term. When both the parameters ( attraction and repulsion ) are non zero 
as in VDW HRG, the normalized entropy density falls in between
the ideal HRG and the EV HRG (i.e $a = 0$). This trend is also true for higher 
chemical potentials as shown in Fig.2b and Fig.2c. The normalized entropy 
density is always large for the large chemical potentials. One can see from Fig.2,
the splitting between different the curves occur at the lower temperature for higher  
chemical potential. As one can see from Fig.2a, the splitting between 
different curves occurs at a temperature $\sim 160$ $MeV$ at $\mu_{B} = 0$. 
This splitting occurs at a lower temperature $\sim 130$ $MeV$ at $\mu_{B} = 0.3$ $GeV$ 
(Fig.2b) and at $\sim 100$ $MeV$ at $\mu_{B} = 0.6$ $GeV$ (Fig.2c). This is because when the 
baryon chemical potential is large, there are more baryons in the system compared to mesons.
Since the VDW interactions have been taken among baryons-baryons and antibaryons-antibaryons,
the VDW parameters plays a significant role at a lower temperature for 
higher chemical potentials. 
          
\begin{figure}
\centering
\subfloat{\includegraphics[scale=0.45]{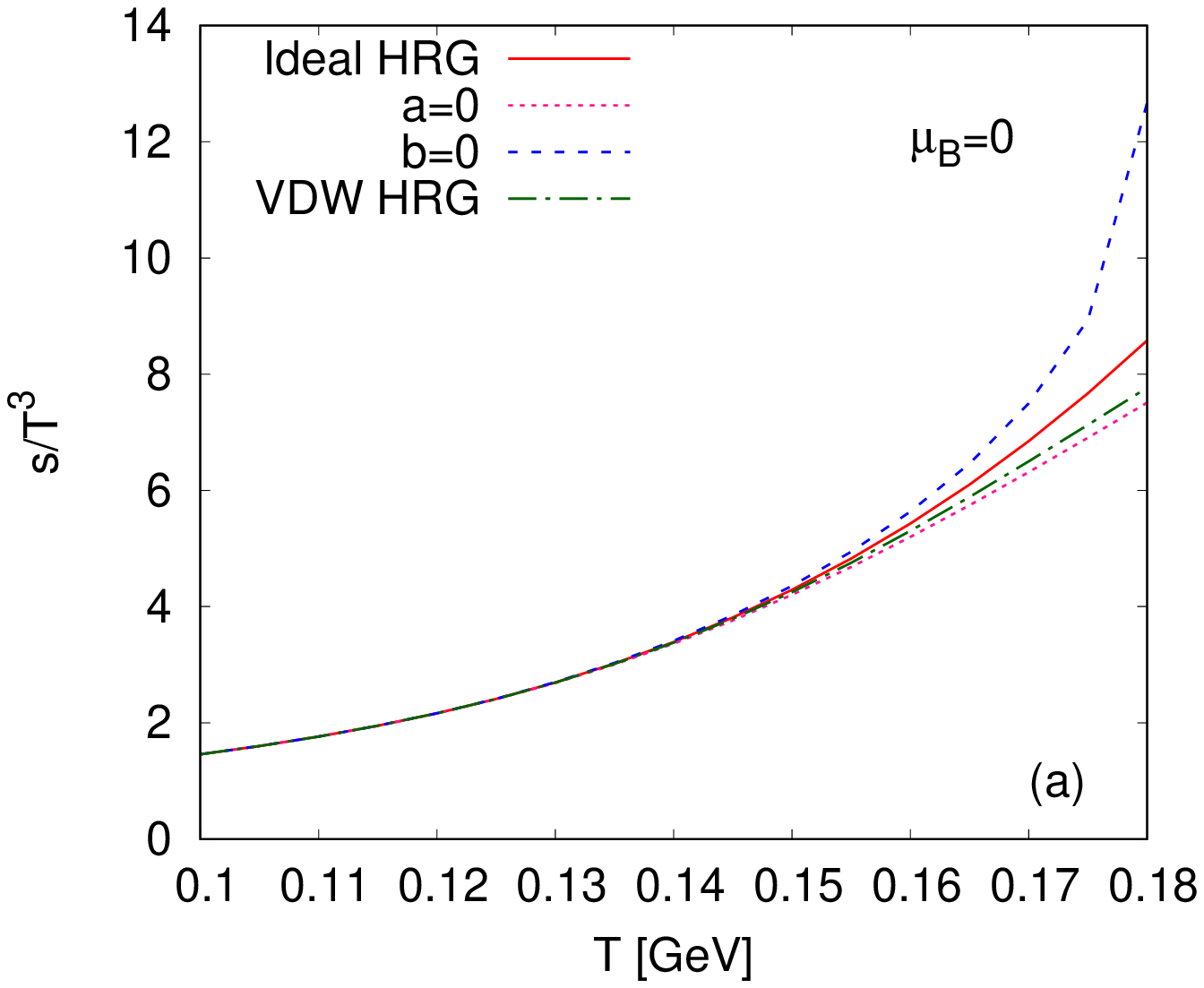}}
\subfloat{\includegraphics[scale=0.45]{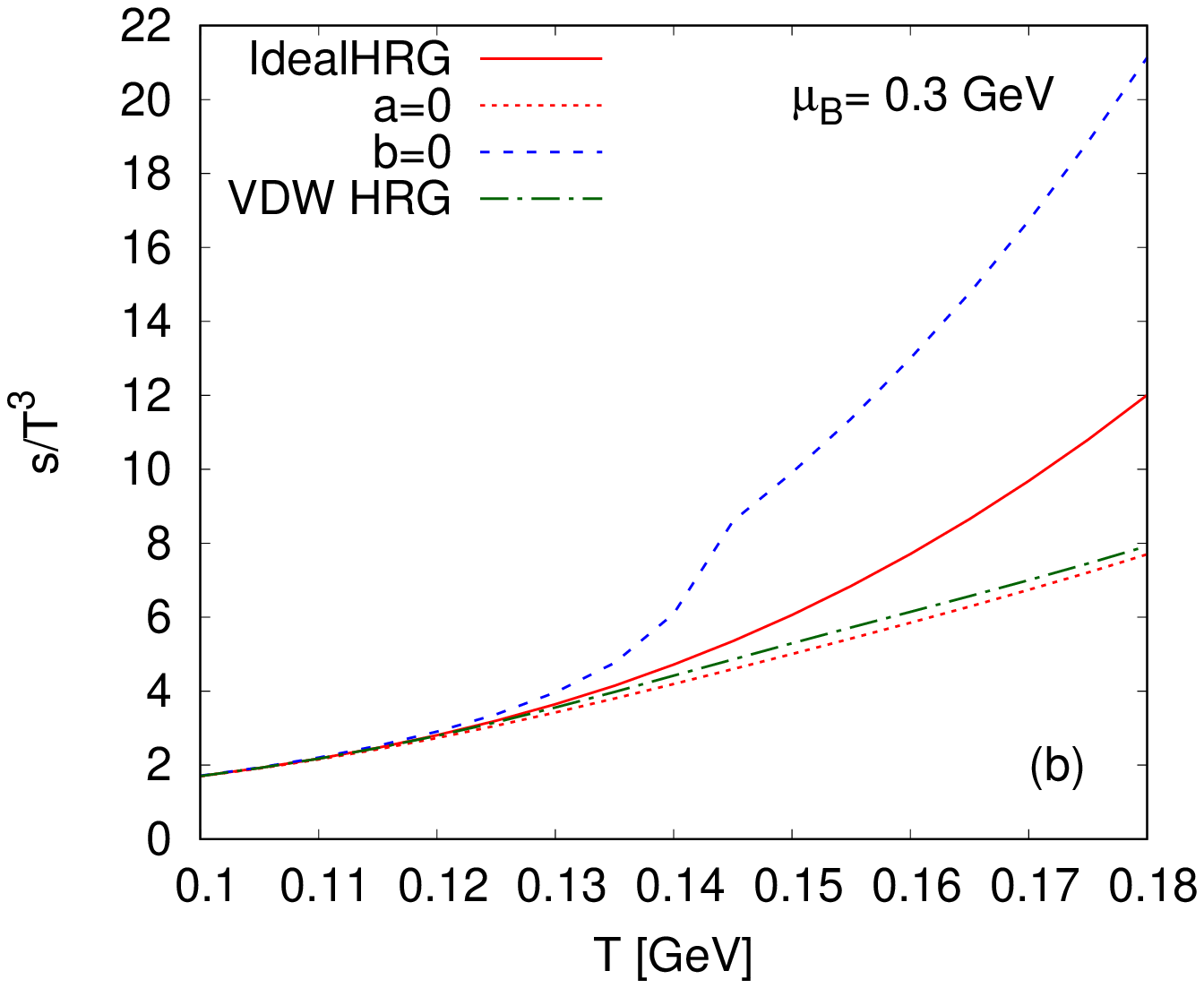}}
\subfloat{\includegraphics[scale=0.45]{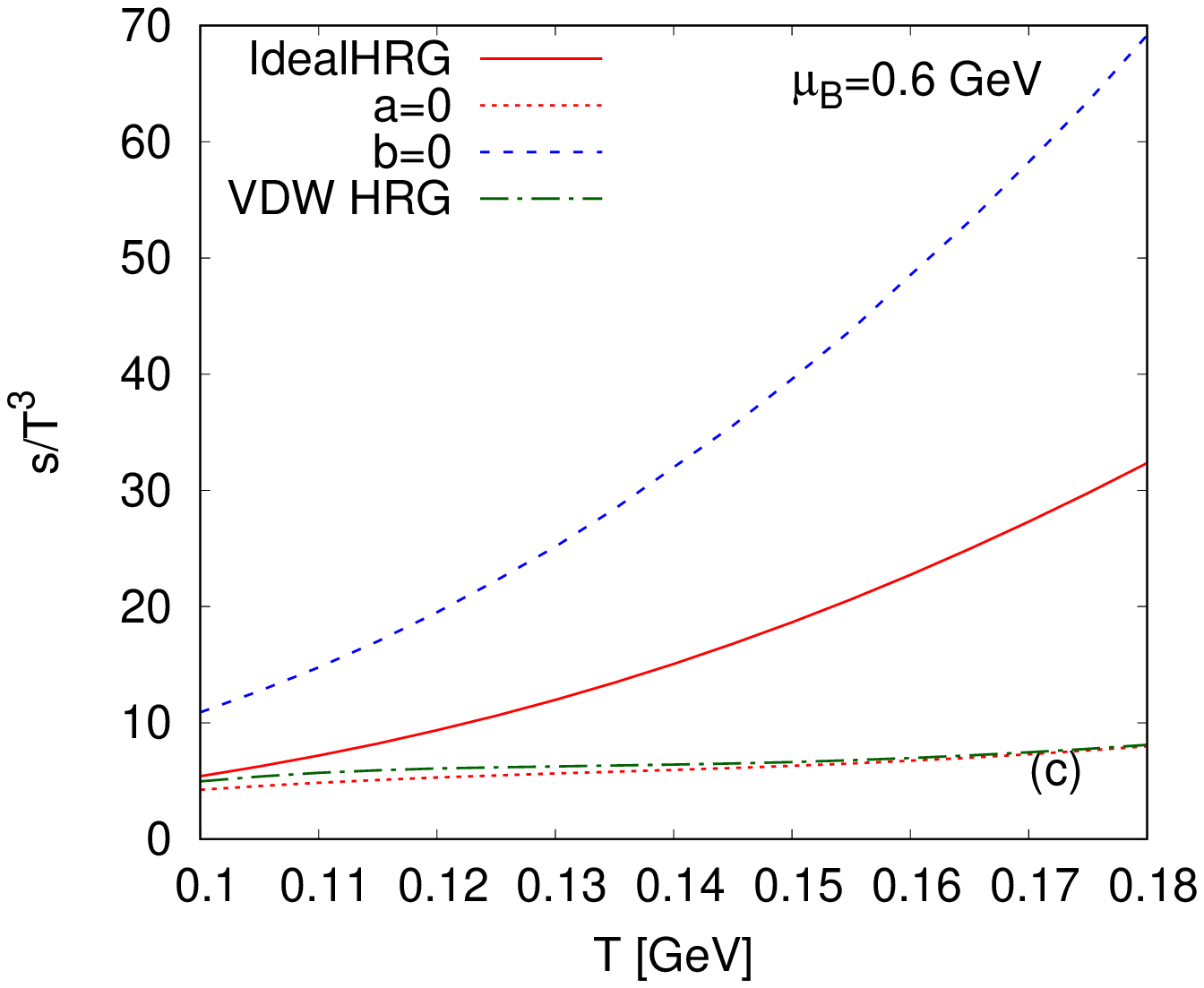}}
\caption{}{The normalized entropy density curve at (a) $\mu_{B} = 0$,
(b) $\mu_{B} = 0.3$ $ GeV$ and (c) $\mu_{B} = 0.6$ $GeV$.}
\label{Fig.2}
\end{figure}

Fig.3 shows the variation of $\eta$ with respect to the temperature at three different 
chemical potentials. $\eta$ increases monotonically with the temperature at any fixed 
chemical potential. This is because the number density increases as the temperature increases.
$\eta$ also increases as the chemical potential increases at
a fixed temperature. This is due to the fact that the number density of baryons increase at 
higher chemical potential.

 \begin{figure*}[!hpt]
\begin{center}
\leavevmode
\epsfysize=6truecm \vbox{\epsfbox{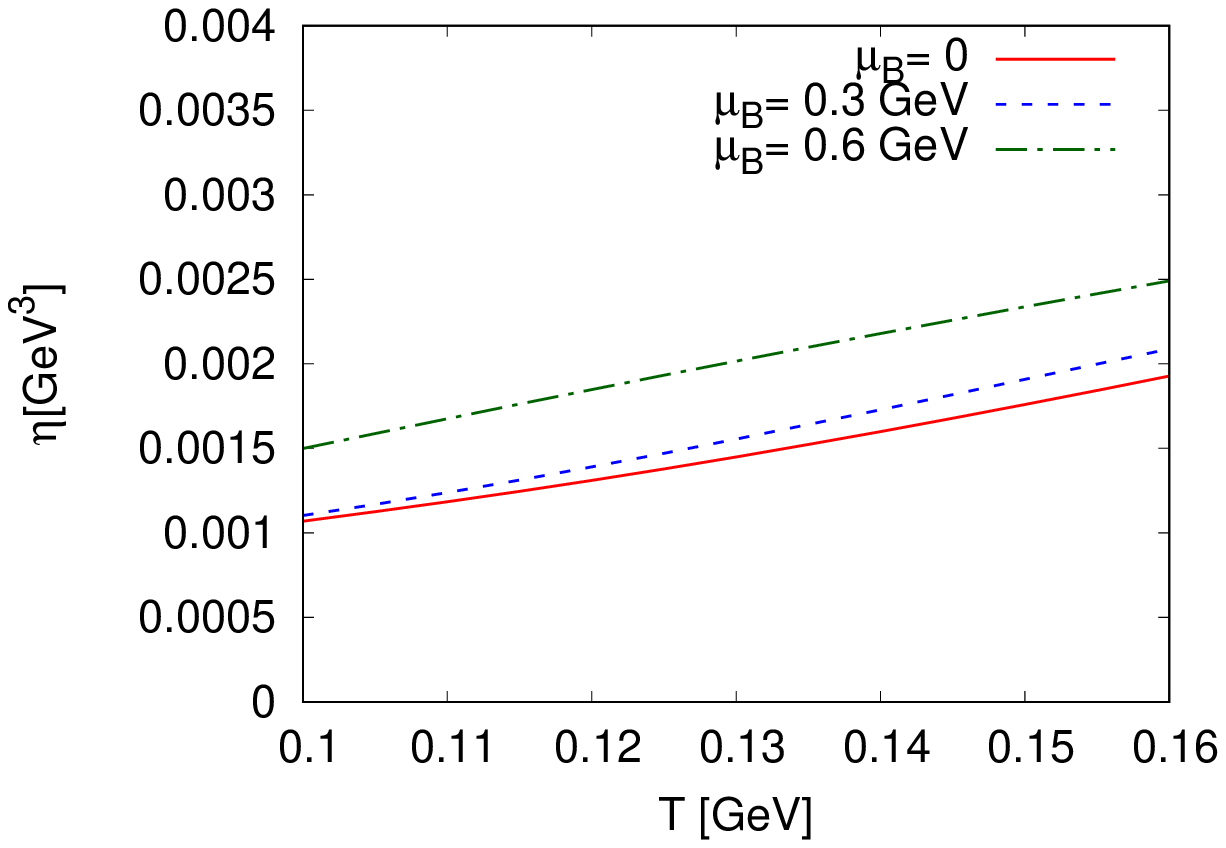}}
\end{center}
\caption{}{Plot of $\eta$ with respect to temperature at different baryon chemical potentials.}
\label{Fig.3}
\end{figure*}

It is very important to measure the dimensionless quantity $\eta/s$ rather than $\eta$.
$\eta/s$ has been shown for two different cases (i.e $a = 0$ and VDW HRG) at 
different chemical potentials in Fig.4.
Fig.4a shows the variation of $\eta/s$ with respect to the temperature at $\mu_{B} = 0$ for
these two different cases. We can see the different curves almost overlap
with each other at $\mu_{B} = 0$. This is because the entropy density is almost 
the same for these two cases upto temperature 160 MeV as shown in Fig.2a. $\eta$ is also
the same for these two different cases since each hadron has the same radius. $\eta/s$
approaches to the famous KSS bound at temperature 160 MeV for the hot hadronic matter
at $\mu_{B} = 0$. This is very close to the critical temperature $\sim 170$ $MeV$   
at $\mu_{B} = 0$ where an almost perfect fluid nature of QGP has been observed.
Fig.4b shows the variation of  $\eta/s$ with respect to the temperature at $\mu_{B} = 300$
$MeV$. $\eta/s$ is almost the same for VDW HRG model and $a = 0$ case due to the 
reasons explained above. This is also true for $\mu_{B} = 600$ $MeV$ as shown   
in Fig.4c. However at higher $\mu_{B} = 600$ $MeV$, $\eta/s$ is different for the two different
cases at lower temperature. This is due to the small difference in the entropy density at lower 
temperature as shown in Fig.2c. One can clearly see $\eta/s$ approaches the KSS bound at a 
relatively higher temperature for higher chemical potential. We also can see $\eta/s$ 
decreases as chemical potential increases at a fixed temperature. This is because 
the entropy density is large for higher chemical potential.   
 
\begin{figure}
\centering
\subfloat{\includegraphics[scale=0.45]{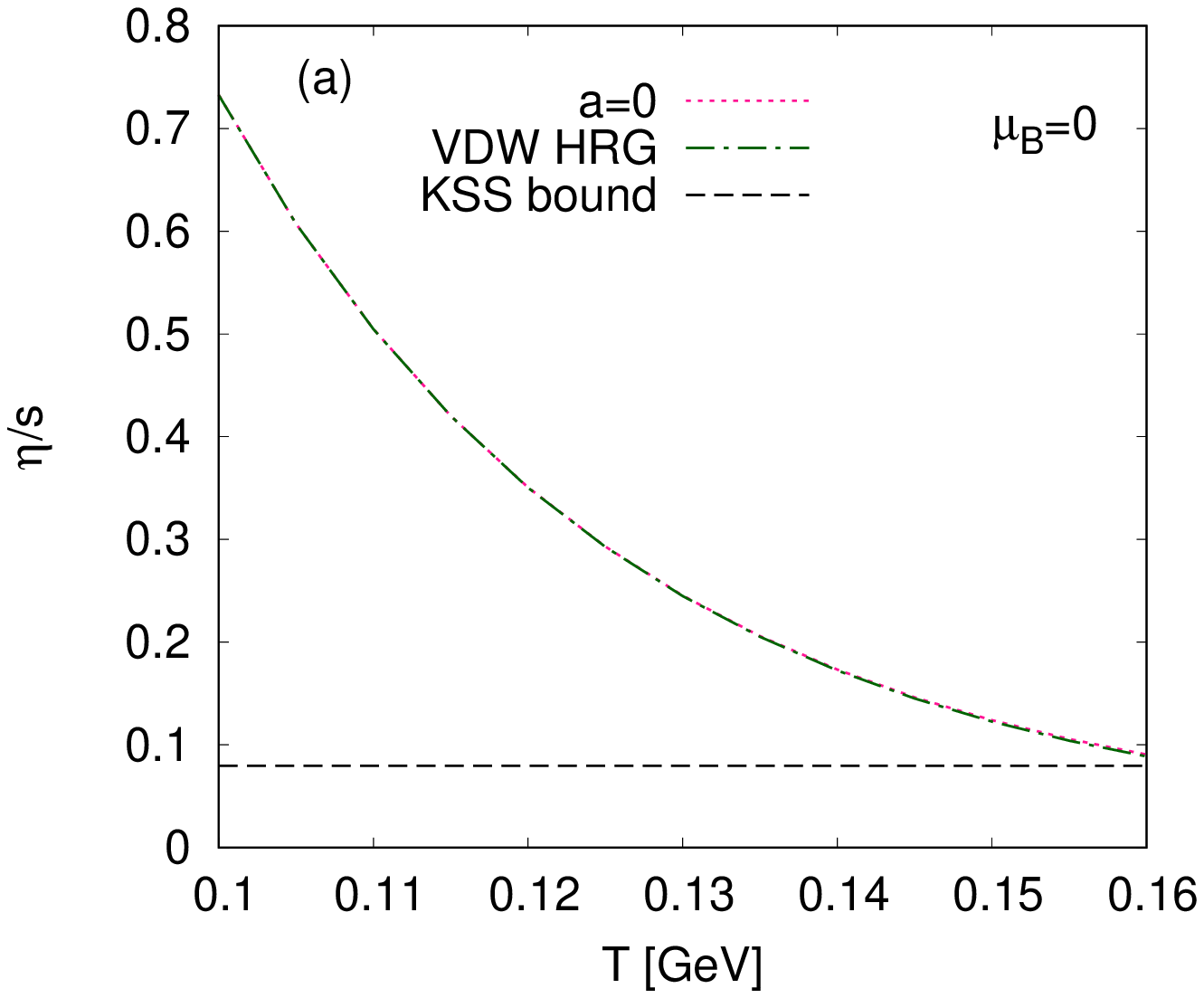}}
\subfloat{\includegraphics[scale=0.45]{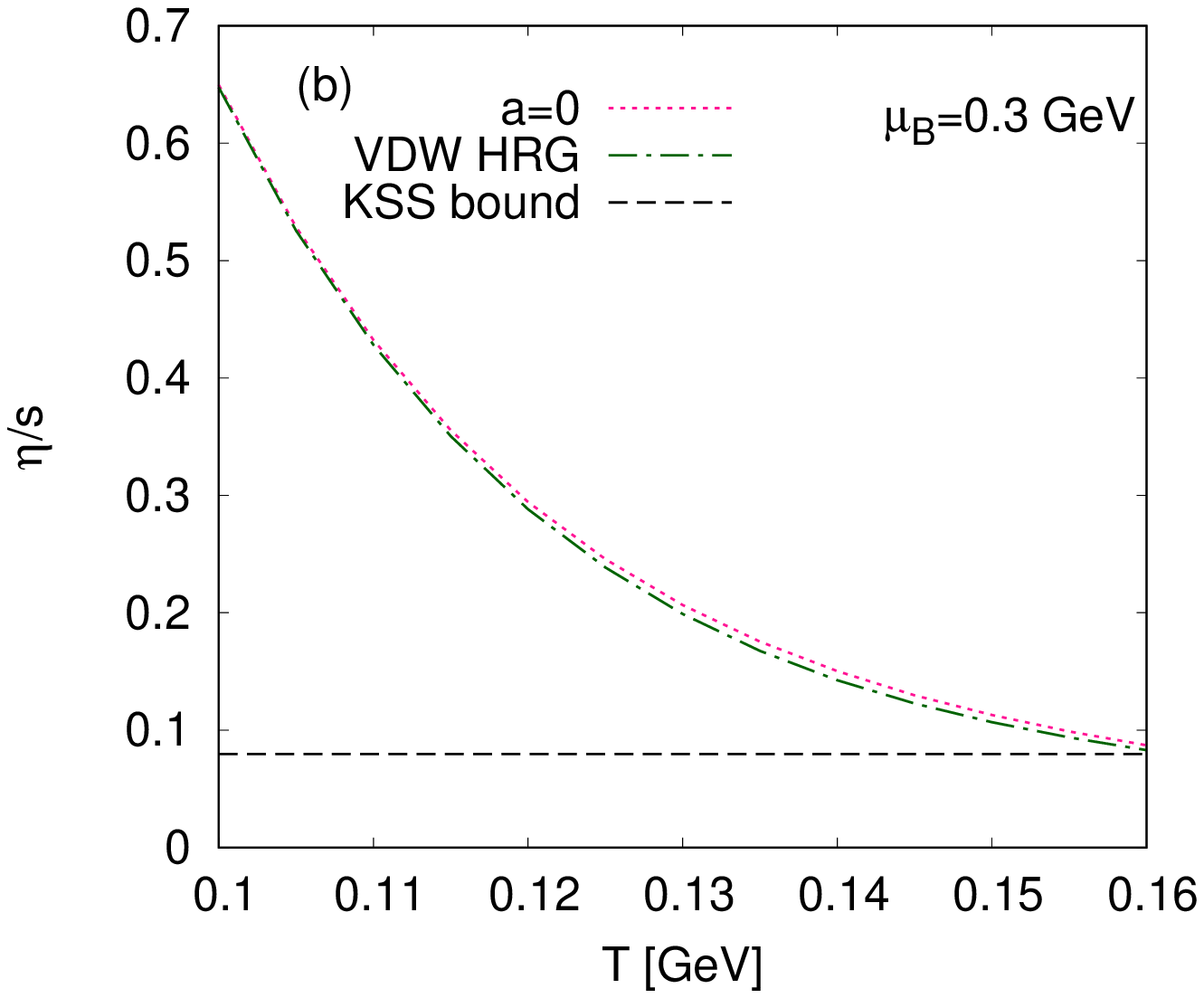}}
\subfloat{\includegraphics[scale=0.45]{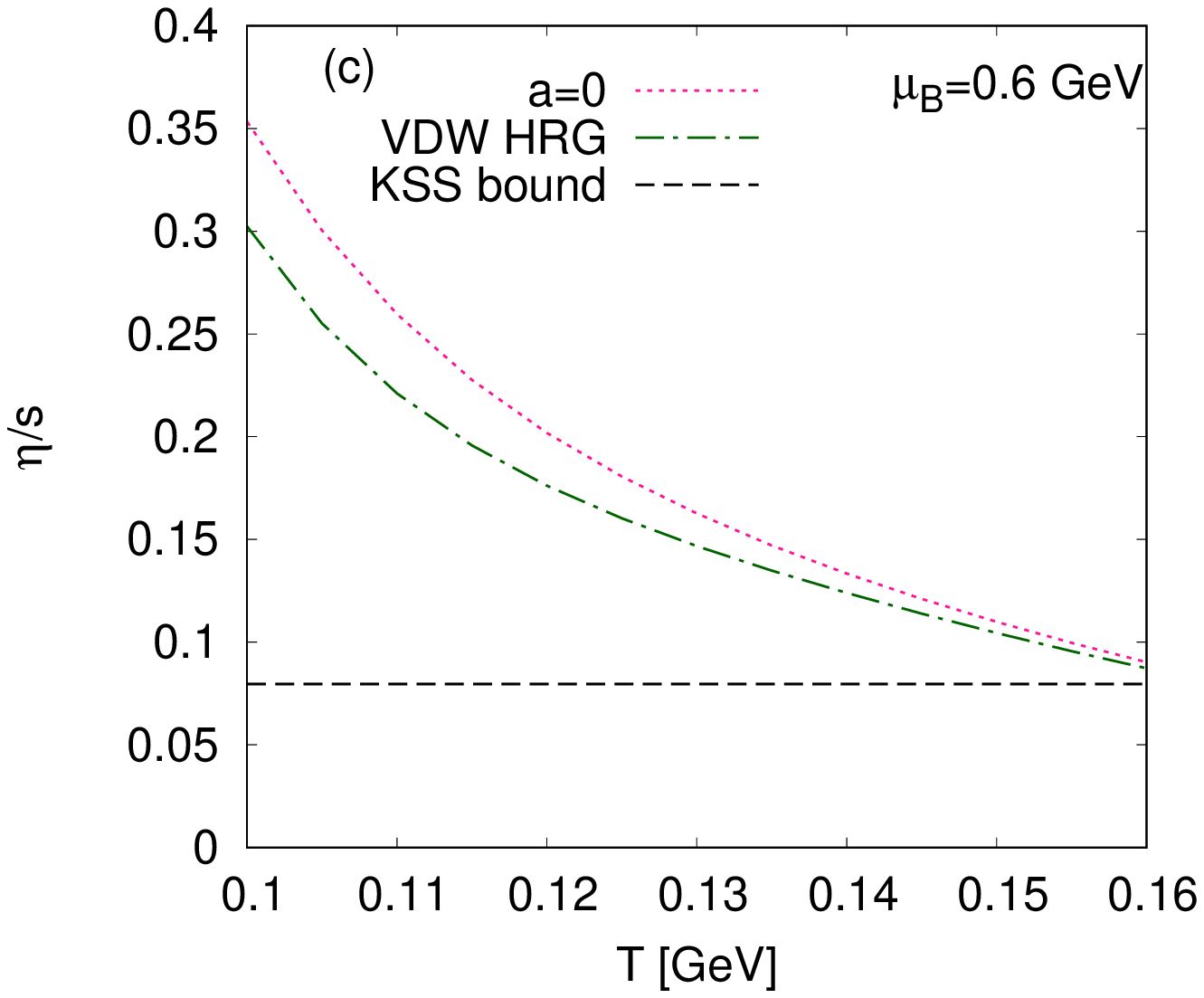}}
\caption{}{Plot of $\eta/s$ with respect to temperature at (a) $\mu_{B} = 0$,
(b) $\mu_{B} = 0.3$ $ GeV$ and (c) $\mu_{B} = 0.6$ $GeV$.}
\label{Fig.4}
\end{figure}

Fig.5 shows the variation of $\zeta$ with respect to temperature at two different chemical
potentials $\mu_{B} = 0$ and $\mu_{B} = 600$ $MeV$. One can see $\zeta$ at $\mu_{B} = 0$ is same 
for both EV HRG and VDW HRG case. This is mainly because the repulsion parameter is same for 
both the case and the attraction parameter in VDW HRG does not affect $\zeta$ at $\mu_{B} = 0$
due to equal number of baryons and antibaryons. When $\mu_{B}$ increases, $\zeta$ increases.
The shear viscosity is larger for VDW HGR compared to EV HRG at $\mu_{B} = 600$ $MeV$.
This is because there are more baryons compared to antibaryons in the system at 
$\mu_{B} = 600$ $MeV$ and the attraction parameter is considered in between baryons- baryons
and antibaryons-antibaryons. 

 \begin{figure}
\leavevmode
\epsfysize=6truecm \vbox{\epsfbox{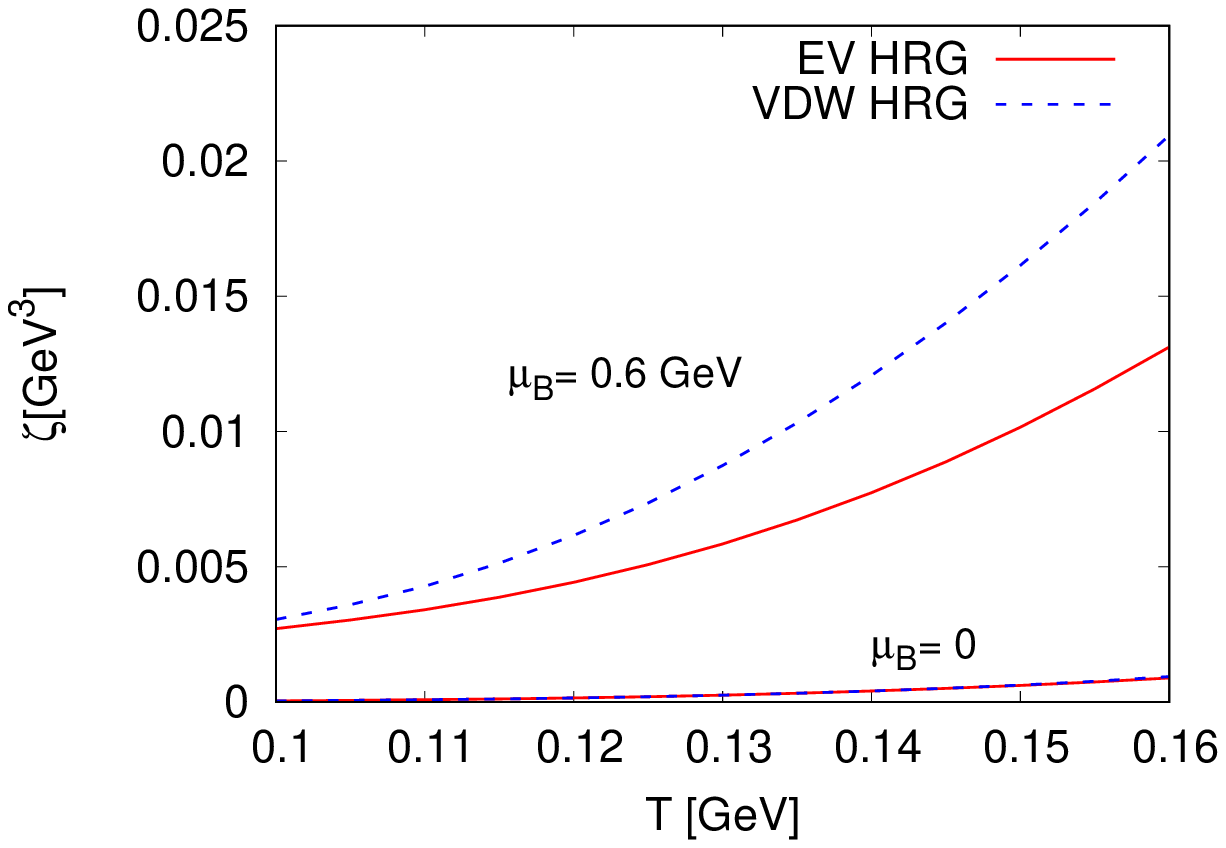}}
\caption{}{Plot of $\zeta$ with respect to temperature at $\mu_{B} = 0$ and $\mu_{B} = 0.6$ $GeV$.}
\label{Fig.5}
\end{figure}

Fig.5 shows the variation of $\zeta/s$ with respect to temperature at two different chemical
potentials $\mu_{B} = 0$ and $\mu_{B} = 600$ $MeV$. $\zeta/s$ is less in VDW HRG compared to 
EV HRG at $\mu_{B} = 0$ due to the fact that the entropy density in VDW HRG is large compared to
EV HRG. This is also true at $\mu_{B} = 600$ $MeV$. The nontrivial variation of $\zeta/s$ with 
temperature in VDW HRG at $\mu_{B} = 600$ $MeV$ is due to the nontrivial variation of 
entropy density.  

\begin{figure}
\epsfysize=6truecm \vbox{\epsfbox{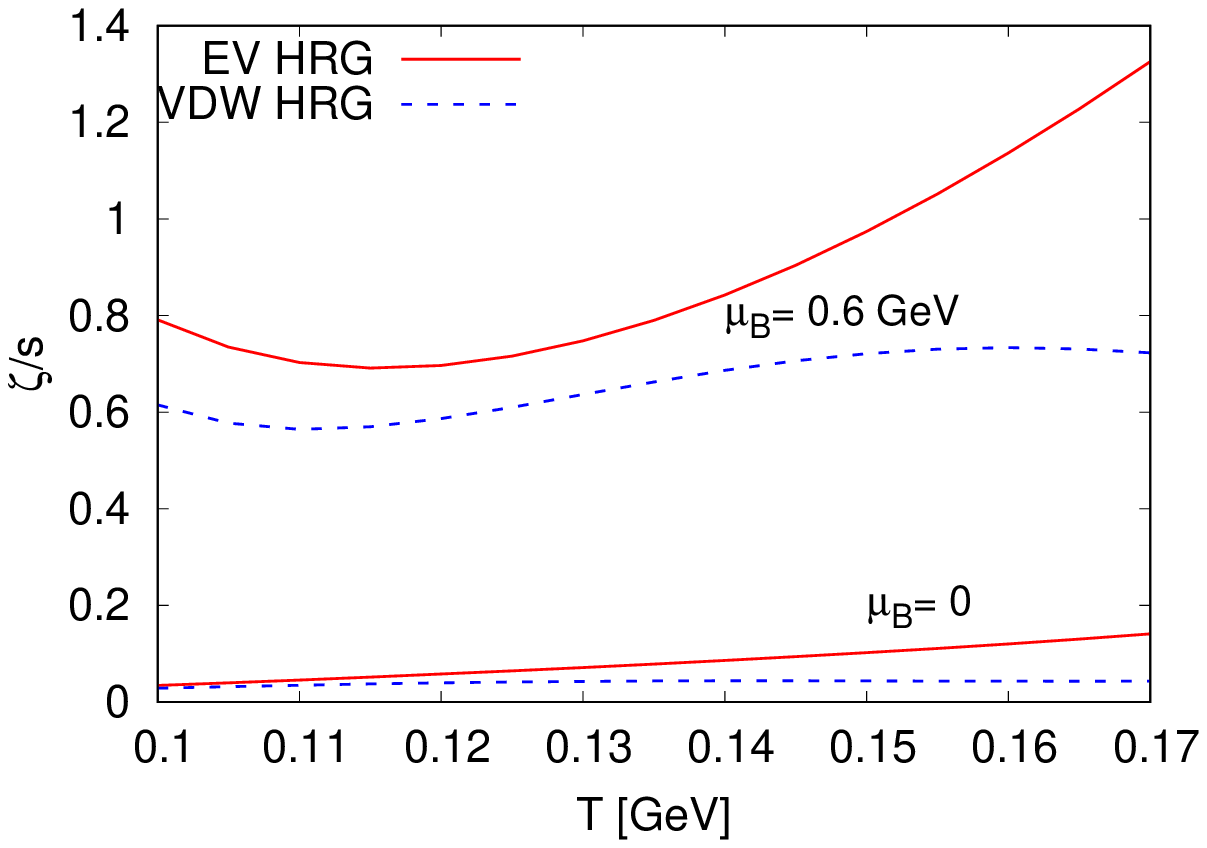}}
\caption{}{Plot of $\zeta/s$ with respect to temperature at (a) $\mu_{B} = 0$ and $\mu_{B} = 0.6$ $GeV$.}
\label{Fig.6}
\end{figure}

Now we estimate the transport coefficients of the hadronic matter along the 
chemical freezeout curve. First we will discuss the effect of VDW attraction and 
repulsion terms on the universal chemical freezeout curve determined by the condition 
$E/N =\epsilon/n \sim 1$ $GeV$ \cite{cleymans}. This is shown in Fig.7. Since $\epsilon$ and $n$ 
are extensive quantities, their ratio is almost independent of the volume. Hence, the
chemical freezeout parameters remain unchanged in the EV HRG model \cite{cleymans2}. However,
the chemical freezeout temperature in VDW HRG is larger than that ideal HRG as
shown in Fig.7. The energy density expression i.e Eq. (\ref{edens}) in VDW HRG has two 
terms. The first term expresses the decrease in energy density due to 
the repulsion parameter $b$ as in the EV HRG model. But the second term (with the
attraction parameter) with a negative sign further decreases the energy
density in VDW HRG model. The number density ( Eq. (\ref{bulkdissi})) in VDW HRG
mainly decreases due to the repulsion term. So, the energy density decreases faster
than the number density in VDW HRG model. So to keep the ratio $\epsilon/n \sim 1$ $GeV$
fixed, the temperature has to be increased in VDW HRG compared to the ideal HRG calculations.
This VDW HRG model with shifted chemical freezeout temperature, the attraction and 
repulsion parameters are $a = 329$~MeV$~$fm$^3$ and $b = 3.42$~fm$^3$ respectively 
along the freezeout line, is described as VDW HRG 1 model.  
It has been shown that there is a shift (decrease) in the chemical freezeout 
temperature at non zero magnetic field due to inverse magnetic catalysis effect
in the HRG model \cite{fuku,ranjita}. Here we show that there is shift (increase) in 
the chemical freezeout temperature due to VDW interactions at zero magnetic field.
       
\begin{figure}
\begin{center}
\leavevmode
\epsfysize=6truecm \vbox{\epsfbox{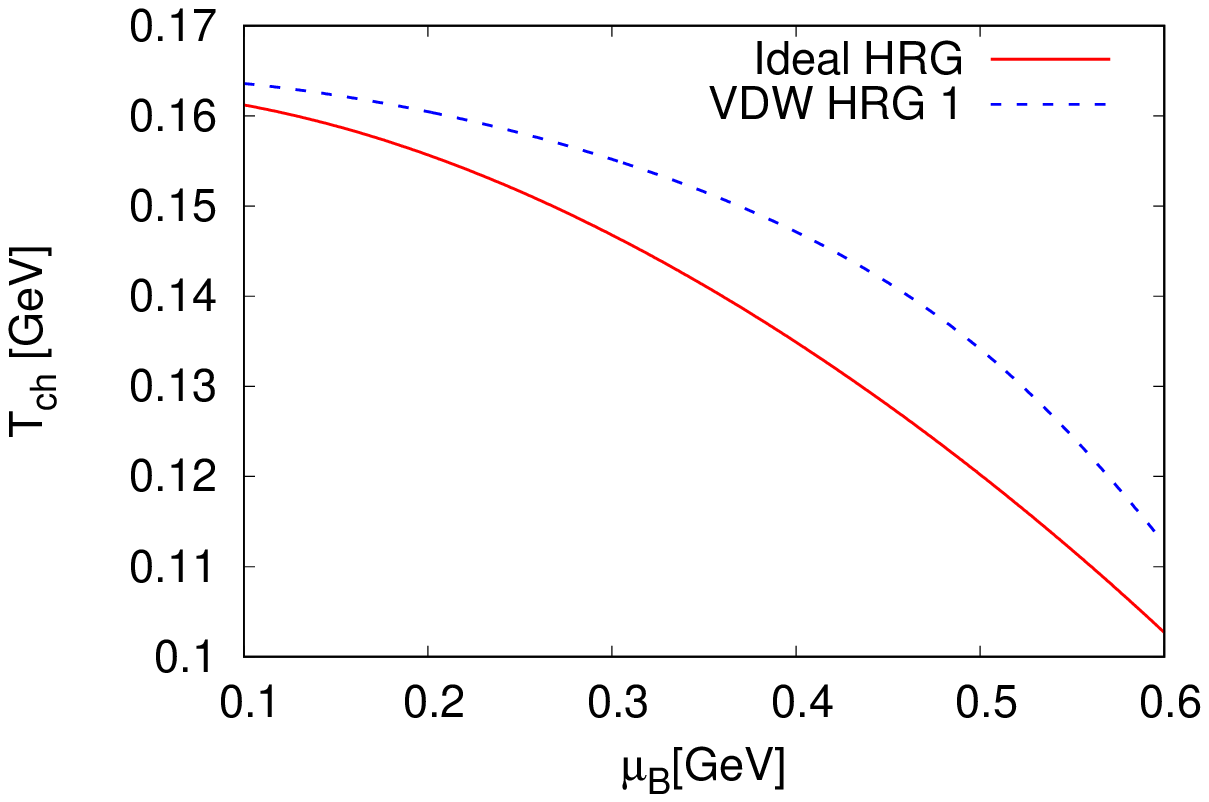}}
\end{center}
\caption{}{The universal chemical freezeout curve determined by $\epsilon/n \sim 1$ $GeV$.}
\label{Fig.7}
\end{figure}

We have taken fixed values of attraction and repulsion parameters i.e 
$a = 329$~MeV$~$fm$^3$ and $b = 3.42$~fm$^3$  at all temperatures
and chemical potentials in all VDW HRG calculations above. However, it has been
already shown that these parameters can vary \cite{nachiketa}. In this work \cite{nachiketa}, 
the authors have taken different values of $a = 926$~MeV$~$fm$^3$ and 
$b = 4.08$~fm$^3$ to match lattice results for $\mu_{B}/T = 1$ and 
$\mu_{B}/T = 2$. Here this value of $b$ corresponds to a larger radius i.e $\sim$ 0.62 fm. 
There is no first principle calculation for these parameters.
Since the repulsion parameter $b$ depends on the radius r of hadron i.e $b = 16\pi r^3/3$, 
we simply can take this to be a fixed parameter i.e $ b = 3.42$~fm$^3$. Once $b$ is fixed and 
the chemical freezeout temperature and baryon chemical potential are fixed 
as for ideal HRG calculations (since these values match with experimental value of 
freezeout parameters) corresponding to the solid curve in Fig.7. 
The universal chemical freezeout curve determined by the condition $\epsilon/n \sim 1$ $GeV$ 
fixes the attraction parameter $a$ as shown in Fig.8. This VDW HRG model with the chemical 
freezeout parameters same as the ideal HRG, the repulsion parameter $ b = 3.42$~fm$^3$ and 
the attraction parameter $a$ varying along the freezeout line is described as VDW HRG 2 model.  
we can see that attraction parameter $a$ increases along the freezeout curve upto 
$\mu_{B} \simeq 300$ $MeV$ and decreases after that. This is because the system is 
meson dominated upto this chemical potential and after this it is baryon 
dominated \cite{cleymans3}. Since VDW interactions have been taken between baryons-baryons, 
antibaryons-antibaryons and switched off between mesons in VDW HRG 2, 
the attraction parameter has to increase with chemical potential initially in the meson 
dominated region and to decrease with chemical potential in the baryon dominated region
to keep the ratio $\epsilon/n \sim 1$ $GeV$ fixed. We also have shown the variation of
attraction parameter $a$ along the freezeout curve for $b = 4.08$~fm$^3$ in Fig.8.      
The energy density and the number density decrease faster corresponding to the large 
radius of hadron for $b = 4.08$~fm$^3$. So to keep the ratio $\epsilon/n \sim 1$ $GeV$ fixed at a 
particular freezeout temperature and chemical potential, the attraction parameter has 
to be increased for  $b = 4.08$~fm$^3$ compared to $ b = 3.42$~fm$^3$ as shown in Fig.8. 
It goes from a meson dominated region to a baryon 
dominated region at a lower chemical potential for higher $b$ as shown in Fig.8.

\begin{figure*}[!hpt]
\begin{center}
\leavevmode
\epsfysize=6truecm \vbox{\epsfbox{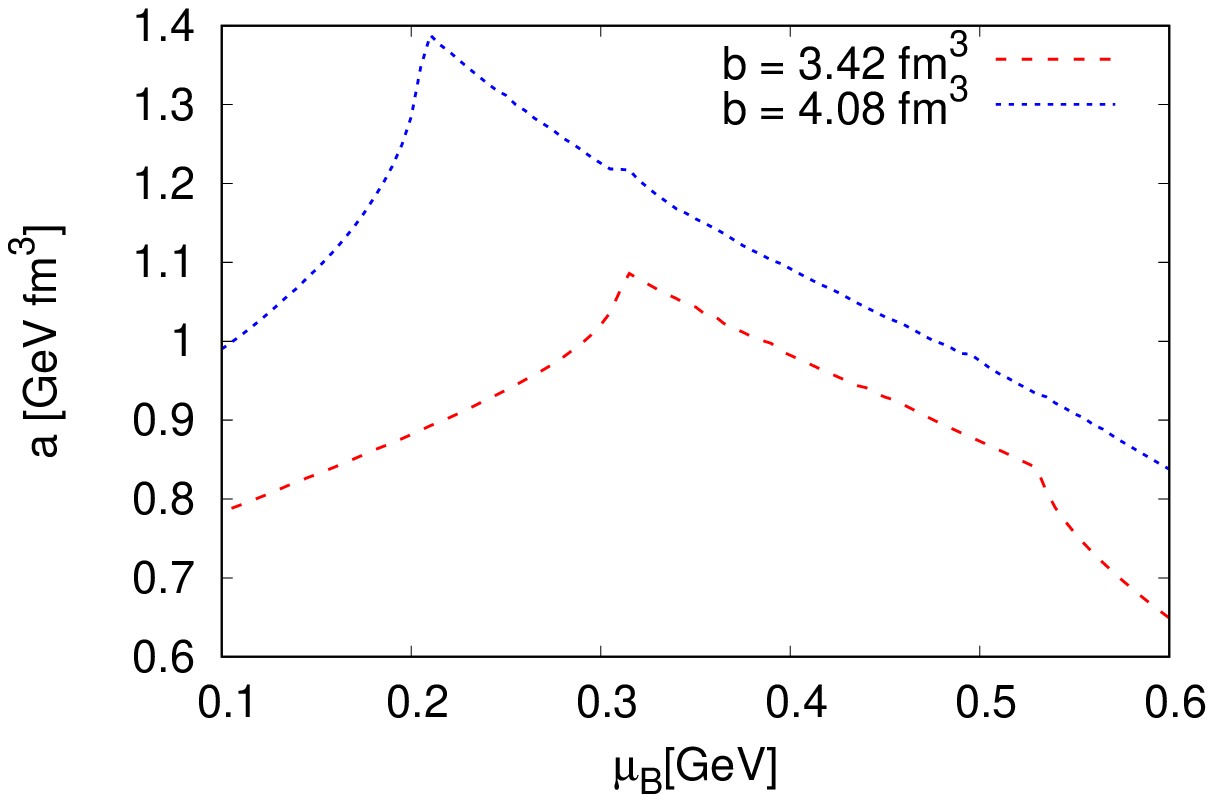}}
\end{center}
\caption{}{Variation of attraction parameter $a$ along the chemical freezeout curve determined 
by $\epsilon/n \sim 1$ $GeV$.}
\label{Fig.8}
\end{figure*} 

We have shown the variation of $\eta/s$ along the freezeout curve for three 
different cases in Fig.9. The dotted line shows  $\eta/s$ for EV HRG case where we have taken
each hadron (all baryons and mesons) has a radius 0.58 fm corresponding to $b = 3.42$~fm$^3$. 
This is along the freezeout 
parameters shown by the solid line in Fig.7. since the chemical freezeout parameters in the 
EV HRG model remain the same as in the ideal HRG. VDW HRG 1 describes $\eta/s$ 
along the shifted chemical freezeout parameters shown by the dashed line in Fig.7. The 
values of the attraction and the repulsion parameters for 
this VDW HRG 1 are $a = 329$~MeV$~$fm$^3$ and $b = 3.42$~fm$^3$ respectively. Fig.9  
shows the variation of $\eta/s$ in VDW HRG 2 along the chemical freezeout curve shown 
by the solid line, where the attraction parameter is varying as shown by the dashed 
curve in Fig.8
and the repulsion parameter $b = 3.42$~fm$^3$ for this case.
Here the repulsion parameter in VDW HRG $b = 3.42$~fm$^3$ corresponds to the same radius
$\sim$ 0.58 fm as in EV HRG. So, $\eta$ is same for EV HRG and VDW HRG 2 case since the chemical 
freezeout parameters and the radius of hadrons are same for both the cases. $\eta$ is relatively
large for VDW HRG 1 due to higher freezeout temperature.  
The entropy density is small in EV HRG model compared to VDW HRG 2 model, since 
there is only repulsion in EV HRG model compared to attractions and repulsions in VDW HRG 2 
model. So, $\eta/s$ in VDW HRG 2 falls below EV HRG model calculations along the freezeout 
curve. Again, the entropy density of VDW HRG 1 is large compared to VDW HRG 2 since the freezeout
temperature is large in VDW HRG 1 compared to VDW HRG 2. So, $\eta/s$ in VDW HRG 1 and VDW HRG 2
are almost the same due to the interplay of large freezeout temperature in VDW HRG 1 and large
attraction parameter in VDW HRG 2. It has been 
shown that $\eta/s$ increases as the chemical 
potential increases along the freezeout line, this is due to the fact that the freezeout
temperature decreases as the chemical potential increases and the entropy density
decreases due the decrease in the freezeout temperature. It is in agreement with the fact that 
at low chemical potentials ( higher collision energies ) QGP is a perfect fluid and the 
hot hadronic matter originating from it respects this nature. 
           
\begin{figure*}[!hpt]
\begin{center}
\leavevmode
\epsfysize=6truecm \vbox{\epsfbox{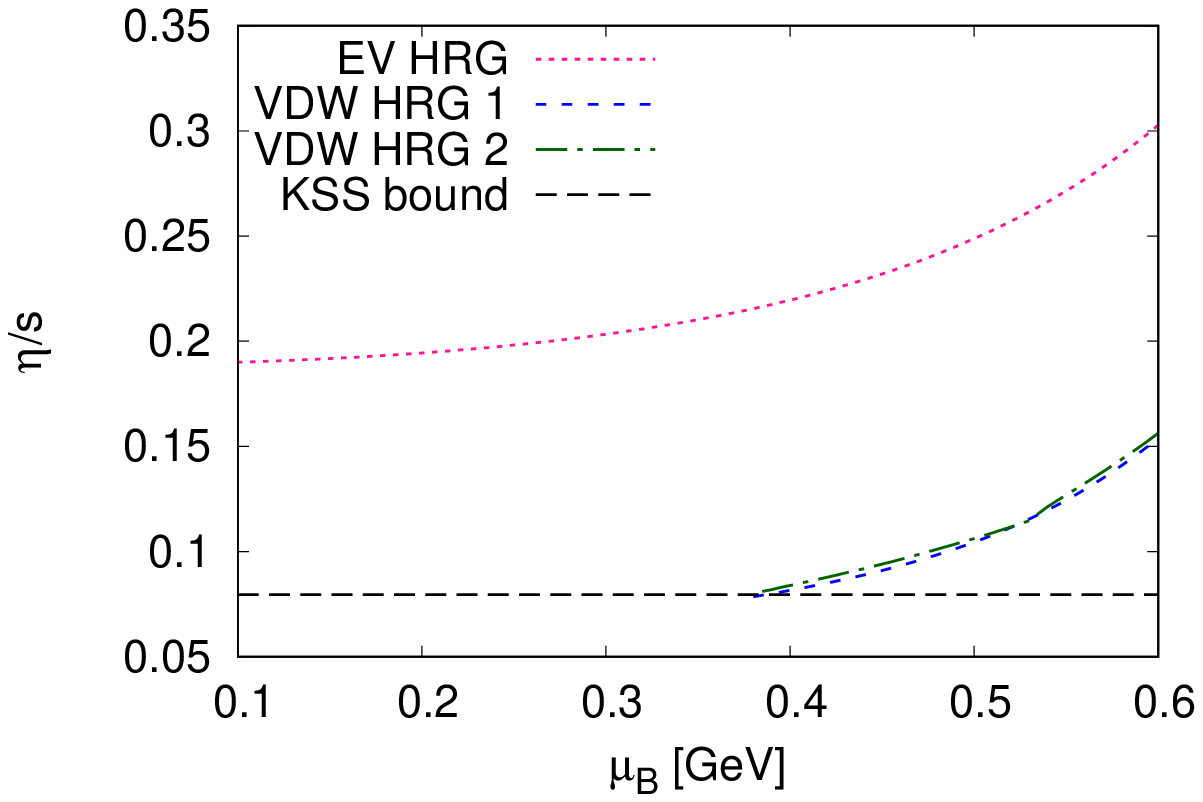}}
\end{center}
\caption{}{Plot of $\eta/s$ along the chemical freezeout curve.}
\label{Fig.9}
\end{figure*}

We have shown the variation of  $\zeta/s$ along the freezeout curve for three different
cases in Fig.10.  $\zeta/s$ increases as the chemical potential increases along the 
freezeout curve due to the decrease in entropy density. $\zeta/s$ is large for EV HRG 
compared to other two cases. The nontrivial dependence of $\zeta/s$ along the freezeout curve
for VDW HRG 2 is due to the nontrivial dependence of attraction parameter along the 
freezeout curve. So the nontrivial dependence of attraction parameter along the freezeout curve
plays an important role in the calculation of bulk viscosity.
 
 \begin{figure*}[!hpt]
\begin{center}
\leavevmode
\epsfysize=6truecm \vbox{\epsfbox{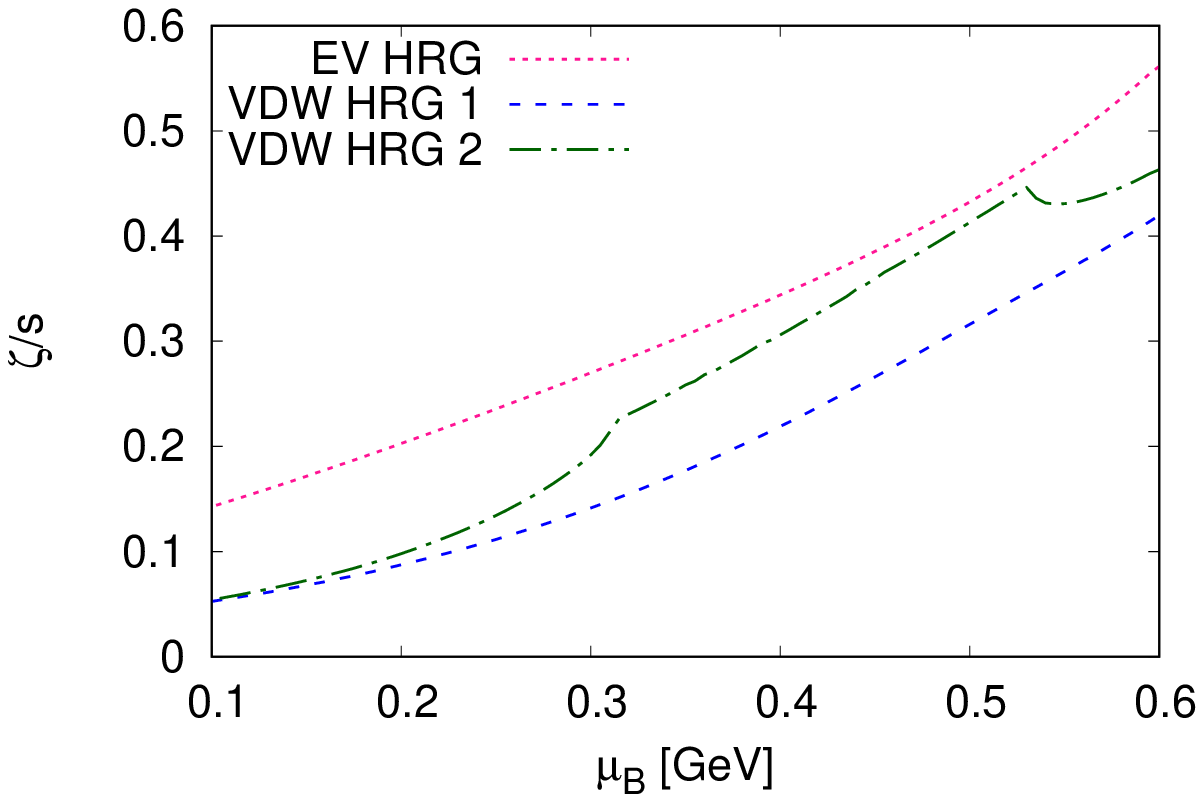}}
\end{center}
\caption{}{Plot of $\zeta/s$ along the chemical freezeout curve.}
\label{Fig.10}
\end{figure*}

\section{CONCLUSIONS}

We have estimated the transport coefficients like shear and bulk viscosity of hot
hadronic matter in VDW HRG model in a relaxation time approximation
and compared these with the estimates obtained in EV HRG model. The average relaxation
time has been calculated with respect to the temperature of the system and this relaxation
time decreases as temperature increases. It has been shown that the attraction and repulsion 
parameters of VDW HRG play an important role in the calculation the 
entropy density at higher $\mu_{B}$ since these VDW interactions has been taken between 
baryons-baryons and antibaryons-antibaryons. The shear viscosity $\eta$ increases as 
the temperature of the system increases and it also increases with the chemical 
potential. But the important quantity $\eta/s$ decreases as temperature increases due to the 
fact that the entropy density increases with respect to temperature. $\eta/s$ for VDW HRG 
is far above the KSS bound at lower temperature and it approaches to the lower bound at higher
temperature. At higher chemical potential, it approaches the KSS bound even at relatively 
higher temperature. The attraction parameter in VDW HRG plays very important role in the 
calculation of bulk viscosity at higher chemical potential. The bulk viscosity is same 
for EV HRG and VDW HRG at $\mu_{B} = 0$, but at higher chemical potential i.e $\mu_{B} = 600$ $MeV$ 
$\zeta$ for VDW HRG is large that EV HRG. However, $\zeta/s$ is larger for EV HRG compared to 
VDW HRG due to the decrease in entropy density in EV HRG compared to VDW HRG. The nontrivial 
dependence of $\zeta/s$ with respect to temperature in VDW HRG at $\mu_{B} = 600$ $MeV$ is 
due to the nontrivial variation of entropy density with respect to temperature in this model.

It is very important to estimate these transport coefficients along the chemical freezeout 
curve where this hadronic system is chemically equilibrated. We have shown the chemical 
freezeout curve determined from the condition $\epsilon/n \sim 1$ $GeV$ for ideal HRG and 
compared it with VDW HRG case. The chemical freezeout curve for VDW HRG is shifted to higher 
temperature due to the fast decrease of the energy density compared to the number density in 
VDW HRG. This chemical freezeout curve with higher freezeout temperature, the attraction
and repulsion parameters as constants along the freezeout curve is represented as 
VDW HRG 1 model. However, one can take the freezeout parameters as in ideal HRG, the repulsion 
parameters corresponding to the hadron radius as fixed quantities. Then the universal freezeout
curve determined from the condition $\epsilon/n \sim 1$ $GeV$ fixes the attraction parameter
along the freezeout curve. This freezeout curve is known as VDW HRG 2 model. The attraction 
parameter in VDW HRG 2 model increases as baryon chemical potential increases upto 300 MeV
and then decreases along the freezeout curve. This is due to the fact that the system is 
meson dominated upto $\mu_{B} = 300$ $MeV$ and then it is baryon dominated. The VDW 
interactions has been taken between baryons-baryons and antibaryons-antibaryons. So the 
attraction parameter is to be increased in the meson dominated region and to be decreased 
in the baryon dominated region to keep $\epsilon/n \sim 1$ $GeV$ fixed at those chemical 
freezeout parameters. We have estimated $\eta/s$ along the freezeout curve for EV HRG, 
VDW HRG 1 and VDW HRG 2 case. $\eta/s$ is larger for EV HRG case compared to other two cases.
This is mainly due to the fact that the entropy density is small in EV HRG model due to only 
repulsion parameter among the hadrons. $\eta/s$ for VDW HRG 1 and VDW HRG 2 are almost same
due to interplay of higher freezeout temperature in VDW HRG 1 and higher attraction parameter
in VDW HRG 2. Similarly,  $\zeta/s$ is larger for EV HRG case compared to other two cases.
 $\zeta/s$ varies nontrivially along the freezeout curve in VDW HRG 2 due to the nontrivial
variation of the attraction parameter. So, these transport coefficients in VDW HRG model 
( with attraction and repulsion among hadrons) are very different than that of EV HRG model
(with only repulsion among hadrons).     

\acknowledgments

We would like to thank Arpan Das and Guru Prakash Kadam for very useful 
discussions and suggestions.



\begin{thebibliography}{99}

\bibitem{paul} P. Romatschke and U. Romatschke, Phys. Rev. Lett. {\bf 99}, 172301 (2007).

\bibitem{alice} K. Aamodt et al. (ALICE Collaboration), Phys. Rev. Lett. {\bf 105}, 252302 (2010).

\bibitem{son} P. Kovtun, D. T. Son, and A. O. Starinets, Phys. Rev. Lett. {\bf 94},
111601 (2005).

\bibitem{lacey} R. A. Lacey, N. N. Ajitanand, J. M. Alexander, P. Chung, W.
G. Holzmann, M. Issah, A. Taranenko, P. Danielewicz, and H.
Stöcker, Phys. Rev. Lett. {\bf 98}, 092301 (2007). 

\bibitem{bazav} A. Bazavov et al., Phys. Rev. {\bf D 90}, 094503 (2014).

\bibitem{bor} S. Borsonyi et al., J. High Energy Phys. {\bf 11} 077 (2010).

\bibitem{arnold} P. Arnold, C. Dogan, and G. D. Moore
Phys. Rev. {\bf D 74}, 085021 (2006).

\bibitem{prl2015} S. Ryu, J.-F. Paquet, C. Shen, G. S. Denicol, B. Schenke, S. Jeon, and C. Gale
Phys. Rev. Lett. {\bf 115}, 132301 (2015).

\bibitem{csernai} L. P. Csernai, J. I. Kapusta, and L. D. McLerran, Phys. Rev. Lett.
{\bf 97}, 152303 (2006).

\bibitem{kapusta1} J. I. Kapusta, in Relativistic Heavy Ion Physics, Landolt-
Bornstein Group 1, Vol. 23 (Springer, New York, 2010).

\bibitem{kharz} D. Kharzeev and K. Tuchin, J. High Energy Phys. {\bf 09},093 (2008).

\bibitem{tuchin} F. Karsch, D. Kharzeev, and K. Tuchin, Phys. Lett. {\bf B 663}, 217 (2008).

\bibitem{plumari} S. Plumari, A. Paglisi, F. Scardina, V. Greco
Phys. Rev. {\bf C 83}, 054902 (2012).

\bibitem{gavin} S. Gavin, Nucl. Phys. {\bf A 435}, 826 (1985).

\bibitem{hm} G. P. Kadam and H. Mishra, Phys. Rev. {\bf C 92}, 035203 (2015).

\bibitem{groot} S. R. de Groot, W. A. van Leeuwen, and C. Weert, Relativistic
Kinetic Theory, Principles and Applications (North-Holland,
Amsterdam, 1980).

\bibitem{marty} R. Marty, E. Bratkovskaya, W. Cassing, J. Aichelin, and H. Berrehrah
Phys. Rev. {\bf C 88}, 045204 (2013).

\bibitem{hm2} A. Abhishek, H. Mishra, and S. Ghosh
Phys. Rev. {\bf D 97}, 014005 (2018).

\bibitem{kapusta2} P. Chakraborty and J. I. Kapusta, Phys. Rev. {\bf C 83}, 014906 (2011). 

\bibitem{bass} N. Demir and S. A. Bass, Phys. Rev. Lett. {\bf 102}, 172302 (2009).

\bibitem{muronga} A. Muronga, Phys. Rev. {\bf C 69}, 044901 (2004).

\bibitem{dubado} A. Dobado and F. J. Llanes-Estrada, Phys. Rev. {\bf D 69}, 116004 (2004).

\bibitem{chen} J.W. Chen, Y. H. Li, Y. F. Liu, and E. Nakano, Phys. Rev. {\bf D 76}, 114011 (2007).

\bibitem{itakura} K. Itakura, O. Morimatsu, and H. Otomo, Phys. Rev. {\bf D 77}, 014014 (2008).

\bibitem{sasaki} C. Sasaki and K. Redlich, Phys. Rev. {\bf C 79}, 055207 (2009).

\bibitem{nachiketa} N. Sarkar and P. Ghosh, Phys. Rev. {\bf C 98}, 014907 (2018).

\bibitem{munzi} P. Braun-Munzinger, K. Redlich and J. Stachel, nucl-th/0304013.

\bibitem{stachel} A. Andronic, P. Braun-Munzinger and J. Stachel, Nucl. Phys. 
{\bf A 772}, 167 (2006).

\bibitem{rischke} D. H. Rischke, M. I. Gorenstein, H. Stocker and W. Greiner, Z. Phys. {\bf C 51}, 
485 (1991).

\bibitem{vdw1} V. Vovchenko, D. V. Anchishkin, and M. I. Gorenstein, J. Phys.
{\bf A 48}, 305001 (2015).

\bibitem{vdw2} V. Vovchenko, D. V. Anchishkin, and M. I. Gorenstein, Phys.
Rev. {\bf C 91}, 064314 (2015).

\bibitem{vdw3} V. Vovchenko, D. V. Anchishkin, M. I. Gorenstein, and R. V.
Poberezhnyuk, Phys. Rev. {\bf C 92}, 054901 (2015).

\bibitem{vdw4} V. Vovchenko, M. I. Gorenstein, and H. Stoecker, Phys. Rev.
Lett. {\bf 118}, 182301 (2017).

\bibitem{pdg2012} Particle Data Group, J. Beringer et al., Phys. Rev. {\bf D 86},
010001 (2012).

\bibitem{cleymans} J. Cleymans and K. Redlich, Phys. Rev. Lett.{\bf 81}, 5284 (1998).

\bibitem{cleymans2} J. Cleymans, H. Oeschler, K. Redlich, and S. Wheaton, Phys.
Lett. {\bf C 73}, 034905 (2006).

\bibitem{fuku} K. Fukushima and Y. Hidaka, Phys. Rev. Lett. {\bf 117}, 102301 (2016).

\bibitem{ranjita} R. K. Mohapatra, arXiv:1711.06913 (Accepted by PRC)

\bibitem{cleymans3} J. Cleymans, arXiv:1711.02882.

\end{thebibliography}
\end{document}